\documentclass[10pt,aps,twocolumn,showpacs,preprintnumbers,amsmath,amssymb,pre]{revtex4-1}
\usepackage{graphicx}
\usepackage{dcolumn}
\usepackage{bm}
\usepackage{amsmath}
\usepackage{epstopdf}
\usepackage{color}
\usepackage{multibib}

\begin{document}

\title{Observation of axisymmetric standard magnetorotational instability in the laboratory}

\author{Yin Wang$^{1}$}
\email{ywang3@pppl.gov}

\author{Erik P. Gilson$^1$, Fatima Ebrahimi$^{1,2}$, Jeremy Goodman$^2$, Hantao Ji$^{1,2}$}
\affiliation{$^1$Princeton Plasma Physics Laboratory, Princeton University, Princeton, New Jersey 08543, USA\\
$^2$Department of Astrophysical Sciences, Princeton University, Princeton, New Jersey 08544, USA}
\begin{abstract}
We report the first direct evidence for the axisymmetric standard magnetorotational instability (SMRI) from a combined experimental and numerical study of a magnetized liquid-metal shear flow in a Taylor-Couette cell with independently rotating and electrically conducting end caps. When a uniform vertical magnetic field $B_i$ is applied along the rotation axis, the measured radial magnetic field $B_r$ on the inner cylinder increases linearly with a small magnetic Reynolds number $Rm$ due to the magnetization of the residue Ekman circulation. Onset of the axisymmetric SMRI is identified from the nonlinear increase of $B_r$ beyond a critical $Rm$ in both experiments and nonlinear numerical simulations. The axisymmetric SMRI exists only at sufficiently large $Rm$ and intermediate $B_i$, a feature consistent with theoretical predictions. Our simulations further show that the axisymmetric SMRI causes the velocity and magnetic fields to contribute an outward flux of axial angular momentum in the bulk region, just as it should in accretion disks.
\end{abstract}
\maketitle

Astronomical accretion disks consist of Keplerian gas or plasma flow about a compact massive object such as a black hole or protostar, and slowly spiraling inward (accreting) by transporting orbital angular momentum outward~\cite{FKR02}.
Being resistant to hydrodynamic turbulence~\cite{JBSG06} and at least partially ionized in the presence of a magnetic field~\cite{EHT21}, most accretion disks probably transport angular momentum by magnetohydrodynamic (MHD) processes, e.g., the magnetorotational instability (MRI)~\cite{velikhov59,chandra60,BH91,BH98} or magnetized winds~\cite{Lesur21}.
Several characteristics in accretion disks are proposed to be related to MRI, including turbulence~\cite{SS73,Pringle81,LL07,SH09}, dynamo mechanism that sustains an ordered component of the magnetic field~\cite{BNST95,HGB96,ROP07,EPS09} and standing waves responsible for the quasiperiodic oscillations in the X-ray light curves~\cite{ABT06,HBFF09}.
To date, almost everything known about astrophysics-related MRI is based on linear analysis or nonlinear numerical simulations, since astronomical observations do not resolve it.

Efforts to find the standard version of MRI (SMRI) most likely to exist in accretion disks are also made in laboratory experiments. SMRI requires the magnetic field parallel to the rotation axis. While some analogs have been tested~\cite{BHP09,Vasil15,BCM15,HBCGJ19}, SMRI remains unconfirmed due to technical challenges and confusions. An earlier claim~\cite{SMTHDHAL04} of the realization of SMRI was due to confusion by Shercliff layer instability~\cite{GJG11}. Related instabilities involving azimuthal fields have been  experimentally demonstrated: helical MRI~\cite{RH05,SGGRSSH06} and azimuthal MRI~\cite{HTR10,SGGGSGRSH14}. Unlike SMRI, these instabilities are inductionless and require an angular velocity profile $\Omega(r)$ steeper than Keplerian, $q=-(r/\Omega)\partial\Omega/\partial r>3/2$, and hence are unlikely to be relevant to most astrophysical disks~\cite{LGHJ06,KSF12}.
Most MRI experiments are conducted in a Taylor-Couette cell consisting of two coaxial cylinders with the gap between the cylinders filled with liquid metal~\cite{SGGRSSH06,NJSRG10} or plasma~\cite{FMEEOPSF20}. For SMRI, the gap must be wide enough to ensure the magnetic diffusion time is longer than the rotation period and the Alfv\'{e}n crossing time~\cite{JGK01,GJ02}.
For infinitely long cylinders of $\partial\Omega/\partial z=0$, $\Omega(r)$ with $q<2$ (``quasi-Keplerian'') has the viscous-driven (``ideal Couette'') form: $\Omega(r)=a+br^{-2}$, which is hydrodynamically linearly stable by Rayleigh's criterion~\cite{Rayleigh17}, yet may still be unstable to SMRI.
Here constants $a$ and $b$ are determined by the cylinder rotations and $(z,r,\phi)$ are cylindrical coordinates.
In experiments, the motion of the end-cap boundaries does not match the ideal Couette profile and Ekman circulation is thus excited, entailing $\partial\Omega/\partial z\ne0$ and (at relevant hydrodynamic Reynolds numbers) some turbulence~\cite{CV66,KJGCS04}.

Here we report an experimental and numerical search for SMRI using a modified Taylor-Couette cell with independently rotating end caps. Ekman circulation is reduced and ideal Couette flow is nearly achieved in its bulk region~\cite{BJSCHLMR06}.
Copper end caps provide inductive coupling to the fluid, enabling the nonlinear saturation of SMRI to detectable levels~\cite{WJGEGJL16,CCEGGJ18,CECGGJ19,WJGEGW20}.
We find that the axisymmetric ($m=0$) SMRI modes, which occur only at sufficiently large rotations and intermediate magnetic field strengths, have been detected for the first time.
Three-dimensional (3D) numerical simulations further show that with SMRI, there is a strong outward flux of axial angular momentum in the bulk region due to both the velocity and magnetic fields, which is similar to that in accretion disks.

\begin{figure}
\centerline{\includegraphics[width=0.35\textwidth]{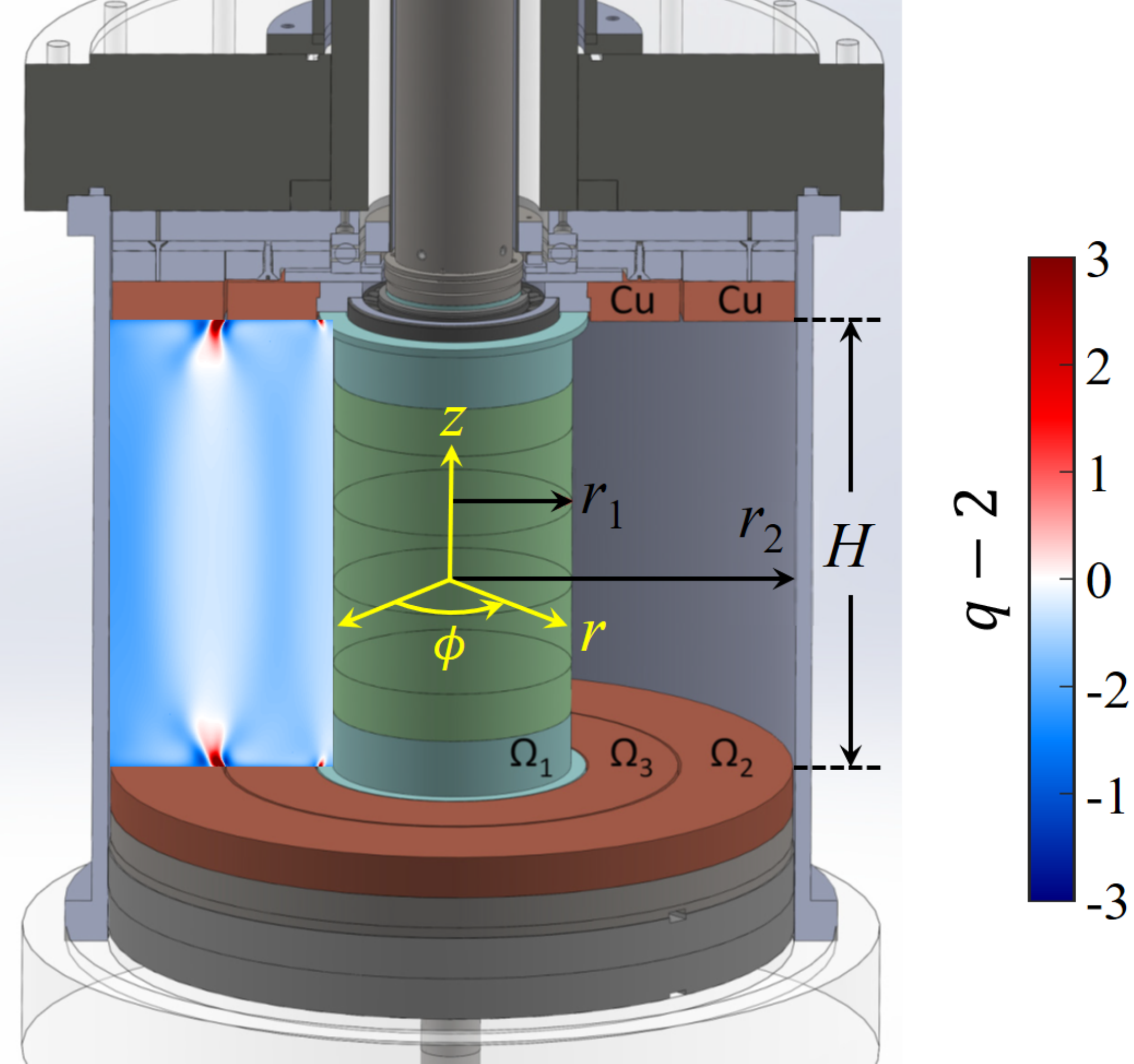}}
\caption{Sketch of the Taylor-Couette cell used in the experiment. The cell has three independently rotatable components: the inner cylinder ($\Omega_1$), the outer-ring-bound outer cylinder ($\Omega_2$), and the upper and lower inner rings ($\Omega_3$). Overlaid on the left is the $\phi$-averaged shear profile, $q-2=-(r/\Omega)\partial\Omega/\partial r-2$, in the statistically steady hydrodynamic state from 3D simulation. The cylindrical coordinate system used is shown in yellow. This plot is adopted from Ref.~\cite{CCEGGJ18}.}
\label{fig1}
\end{figure}

Details about the experimental setup have been described elsewhere~\cite{CCEGGJ18}, and here we only mention some key points. The radii of the inner and outer cylinders (Fig.~\ref{fig1}) are $r_1=7.06$ cm and $r_2=20.3$~cm. The cylinder height is $H=28.0$~cm, giving an aspect ratio $\Gamma\equiv H/(r_2-r_1)\simeq2.1$.
The inner cylinder is composed of five insulating Delrin rings (green) and two caps (cyan) made of stainless steel (conductivity $\sigma_s=1.45\times10^6$~Ohm$^{-1}$m$^{-1}$). These steel caps have a 1~cm protruding rims, which help to further reduce the Ekman circulation~\cite{EJ14}.
The outer cylinder is made of stainless steel.
A GaInSn eutectic alloy (galinstan) (67\% Ga, 20.5\% In, 12.5\% Sn, density $\rho_g=6.36\times10^3$~kg/m$^3$, conductivity $\sigma_g=3.1\times10^6$~Ohm$^{-1}$m$^{-1}$), liquid at room temperature, is used as the working fluid.
The angular velocities of the inner and outer cylinders are $\Omega_1$ and $\Omega_2$, respectively.
Both the upper and lower end caps are made of copper (conductivity $\sigma_c=6.0\times10^7$~Ohm$^{-1}$m$^{-1}$) and split into two rings at $r_3=13.5$~cm.
The inner rings rotate independently at $\Omega_3$. The outer rings are fixed to the outer cylinder and rotate at $\Omega_2$.
For results herein, $\Omega_1:\Omega_2:\Omega_3=1:0.19:0.58$.
Six coils provide a uniform axial magnetic field $B_i \leq 4800$~G.
Hall probes on the inner cylinder with various azimuths at $z/H=0.25$ measure the time series of the local radial magnetic field $B_r(t)$.
Dimensionless measures of the rotation and field strength are the magnetic Reynolds number $Rm=r_1^2\Omega_1/\eta$ and the Lehnert number $B_0=B_i/(r_1\Omega_1\sqrt{\mu_0\rho_g})$, which are varied in the ranges $0.5\lesssim R_m \lesssim4.5$ and $0.05\lesssim B_0\lesssim1.2$. Here $\mu_0$ is the vacuum permeability, $\nu$ and $\eta$ are the kinematic viscosity and magnetic diffusivity of galinstan, and the magnetic Prandtl number $P_m=\nu/\eta=1.2\times10^{-6}$.
The device spins up for 2 minutes, which is several times the Ekman spin-up time (approximately 40 seconds)~\cite{KJGCS04}, thus ensuring a statistically steady flow before the introduction of $B_i$.
When $B_i$ is applied, the flow relaxes a new MHD state that is statistically steady within 2 seconds.

Our 3D simulation uses the code SFEMaNS, which solves the coupled Maxwell and Navier-Stokes equations using spectral and finite-element methods in a fluid-solid-vacuum domain modeled on our experiment~\cite{GLLN09}.
Its main difference from the experiment is that the Reynolds number $Re=r_1^2\Omega_1/\nu=10^3$, versus $Re\sim10^6$ in the experiment.
The simulation also has two stages: it is first run without a magnetic field to a statistically steady hydrodynamic state, followed by the imposition of $B_i$ and continuing until the MHD state saturates.
Plots in Fig.~\ref{fig1} show that except in small regions adjacent to the end-cap ring gaps, $q<2$, indicating hydrodynamically stable (quasi-Keplerian) flow, as desired. Other details of numerical simulations are given in the appendix.

\begin{figure}
\centerline{\includegraphics[width=0.48\textwidth]{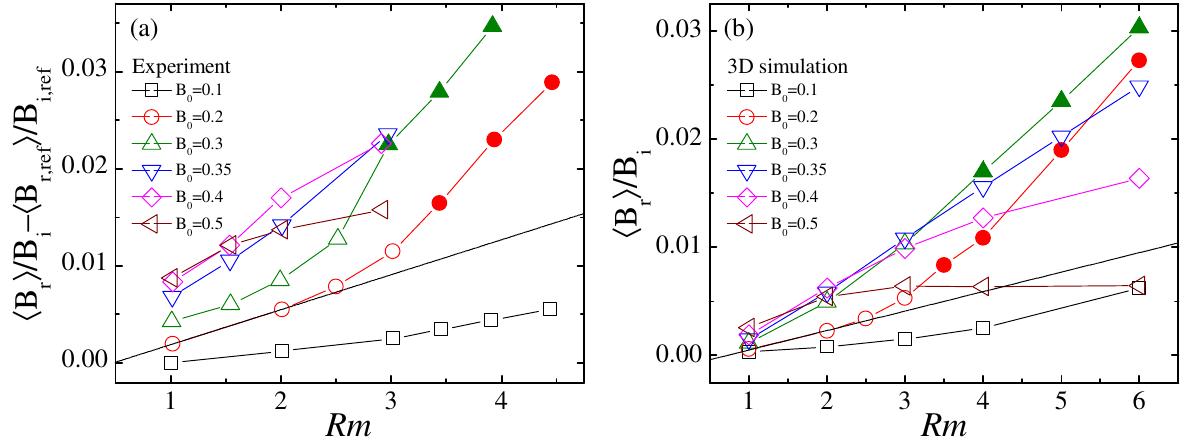}}
\caption{Measured normalized radial magnetic field $\langle B_r\rangle/B_i$ from experiment~(a) and simulation~(b), as a function of $Rm$ for different values of $B_0$. The measurements are conducted at the inner cylinder with $z/H=0.25$. The experimental data include an offset subtraction of $\langle B_{r,\text{ref}}\rangle/B_{i,\text{ref}}$, which does not apply to the simulation data.
The solid lines indicates linear fits to data points at $Rm\lesssim2.5$ and $B_0=0.2$. Solid (open) symbols represent cases with (without) prominent axisymmetric SMRI, defined as $\psi\gtrsim0.005$ according to Eq.~(\ref{eq1}).}
\label{fig2}
\end{figure}

Figure~\ref{fig2}a shows the experimentally measured normalized radial magnetic field, $\langle B_r\rangle/B_i-\langle B_{r,\rm{ref}}\rangle/B_{i,\rm{ref}}$ as functions of $Rm$, in the saturated MHD state. Here $\langle...\rangle$ represents averages over both time and azimuth. $B_{r,\rm{ref}}$ and $B_{i,\rm{ref}}$ are the measured radial magnetic field and the imposed axial magnetic field at the same $Rm$ as $\langle B_r\rangle$ but with $B_0=0.05$. Since a weak magnetic field acts as a passive tracer for the hydrodynamic flow, the non-zero offset $\langle B_{r,\rm{ref}}\rangle/B_{i,\rm{ref}}$ is mainly contributed by experimental imperfections, such as a small axial component of the Hall probe pointing.
After the subtraction of $\langle B_{r,\rm{ref}}\rangle/B_{i,\rm{ref}}$, $\langle B_r\rangle/B_i$ represents the mean radial magnetic field induced only by MHD effects.
As illustrated by the solid line, for small $Rm$, $\langle B_r\rangle/B_i-\langle B_{r,\rm{ref}}\rangle/B_{i,\rm{ref}}$ at different $B_0$ increase almost linearly with $Rm$.
This is caused by the magnetization of the residual Ekman circulation in the hydrodynamic state, which breaks the translational symmetry in the $z$-direction compared to the case of infinitely long cylinders~\cite{GS86}. The onset of axisymmetric SMRI at larger $Rm$ thus can be viewed as an imperfect bifurcation~\cite{GGJ12}, manifesting as an accelerated increase (a ``knee'' in the slope) with $Rm$ in the order parameter such as the radial magnetic field~\cite{WJGEGJL16,WJGEGW20}. This ``knee'' exists only for $0.2\lesssim B_0\lesssim 0.35$ and we infer that SMRI is prominent for solid data points that lie significantly above the linear extrapolation from lower $Rm$.
The ``knee" does not exist outside of this intermediate range for $B_0$. At $B_0=0.1$, the linear increase with $Rm$ continues to the highest $Rm=4.5$ achievable in the experiment. For $B_0\gtrsim0.35$, the increase even becomes slower at larger $Rm$, and the lack of data points at $Rm\gtrsim0.35$ is due to the power limitation of the motor controlling $\Omega_2$.
Figure~\ref{fig2}b shows that the $\langle B_r\rangle/B_i$ obtained from simulation has a similar $Rm$-dependence as the experiment, i.e., a ``knee'' exists only for $0.2\lesssim B_0\lesssim 0.35$ and disappears at weaker or stronger $B_0$.

\begin{figure}
\centerline{\includegraphics[width=0.48\textwidth]{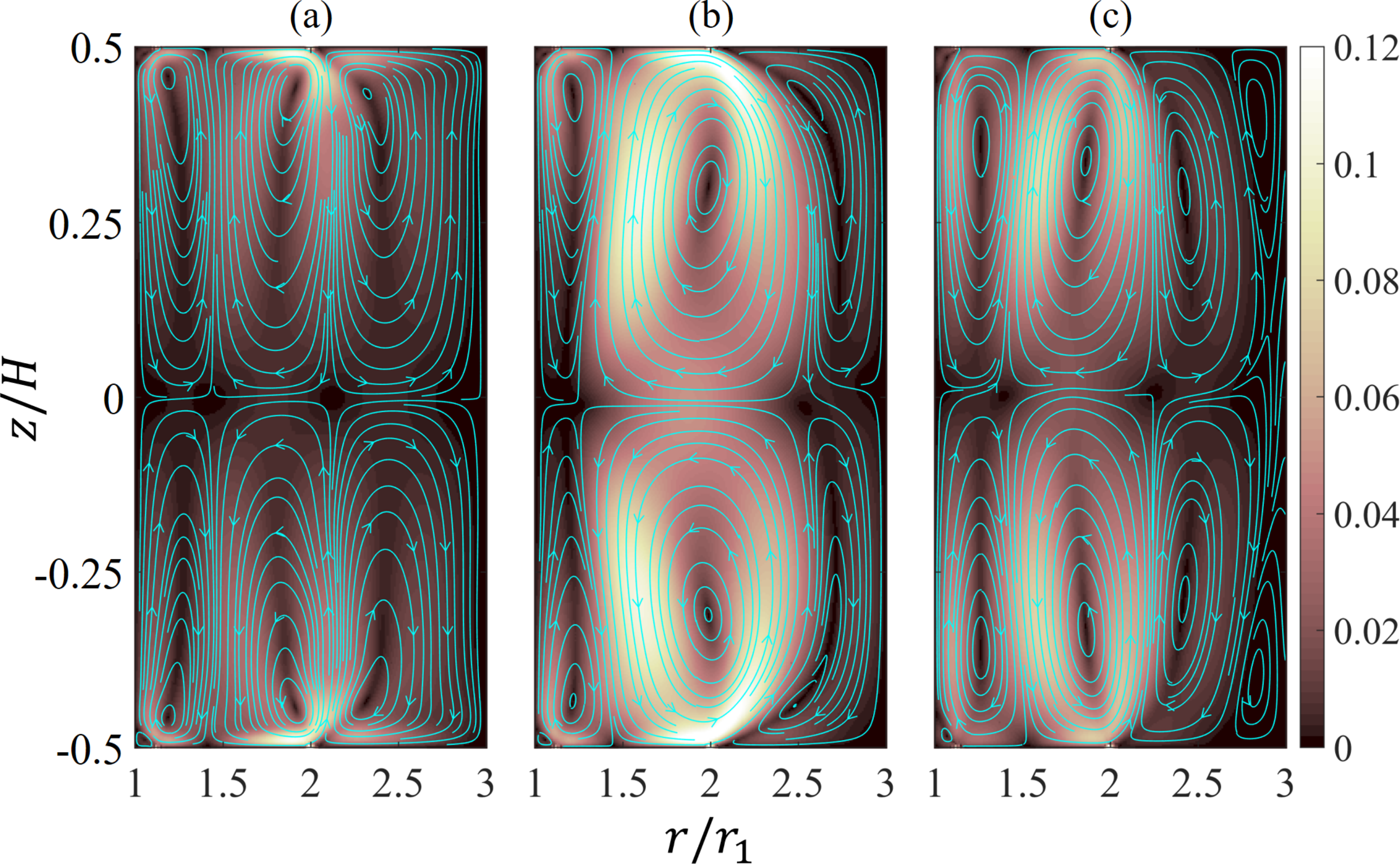}}
\caption{Time- and $\phi$-averaged normalized meridional speed $\sqrt{(u_r^2+u_z^2)}/(r_1\Omega_1)$ from 3D simulations. Streamlines are shown as cyan curves. The data were obtained at $Rm=6$ and different values of $B_0$, with (a) $B_0=0$ (hydrodynamic), (b) $B_0=0.2$ (SMRI), and (c) $B_0=0.5$.}
\label{fig3}
\end{figure}

Simulations further reveal (Fig.~\ref{fig3}) that for $Rm=6$, the axisymmetric SMRI at $B_0=0.2$ results in a significant enhancement of the two vertically stacked circulations in the middle of the $r-z$ plane in the hydrodynamic state ($B_0=0$), as also observed in two-dimensional (2D) simulations~\cite{WJGEGW20}. This enhancement is greatly reduced at $B_0=0.5$, where SMRI is suppressed by an excessive magnetic field. Meanwhile, at $B_0=0.2$ the enhancement becomes prominent only for $Rm\gtrsim4$ [see Fig.~S1 in Supplemental Material (SM)], consistent with the requirement ($Rm\gtrsim3.5$) for SMRI shown in Fig.~\ref{fig2}b.

\begin{figure}
\centerline{\includegraphics[width=0.48\textwidth]{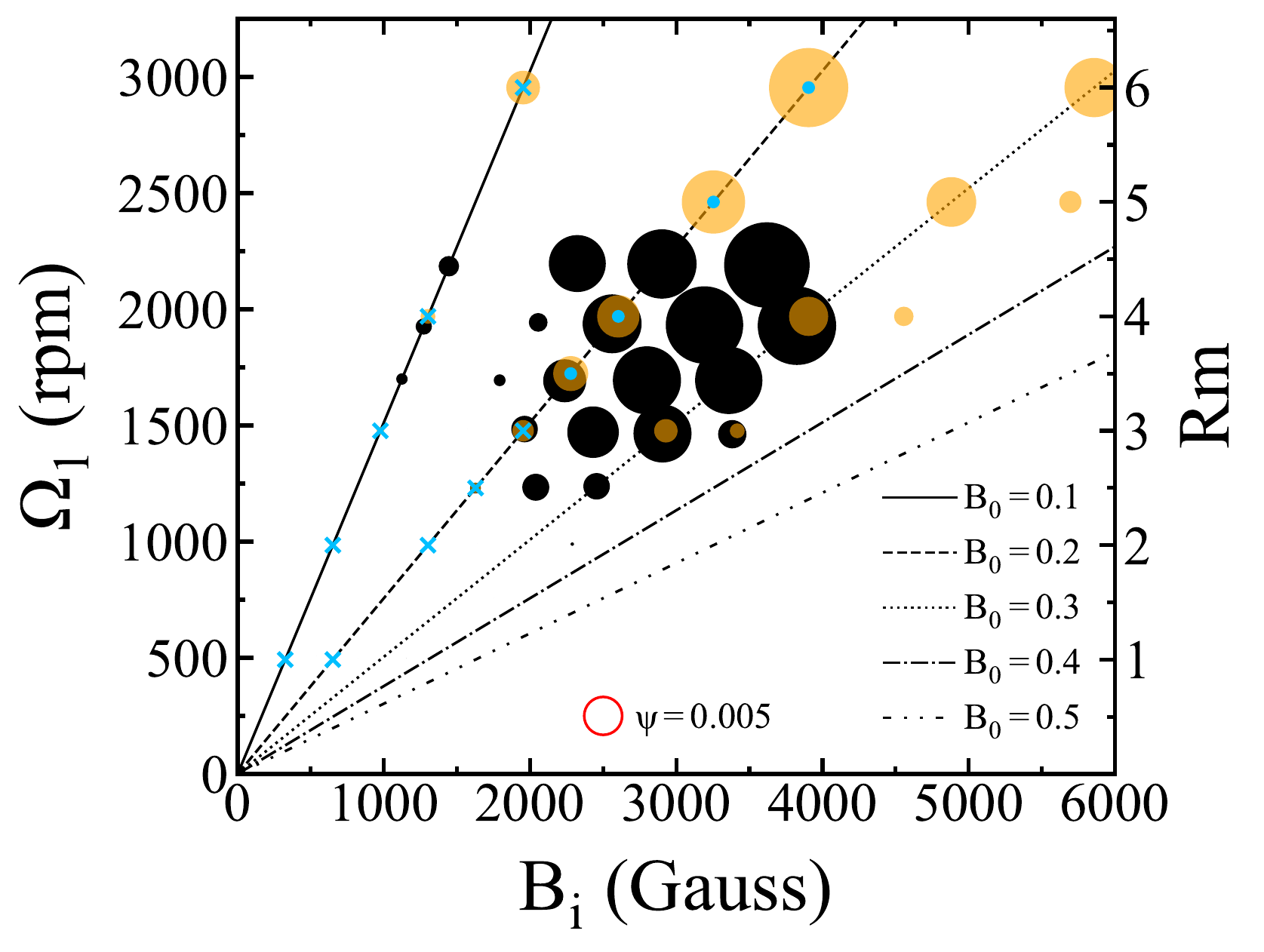}}
\caption{Bubble plot of $\psi$ defined in Eq.~(\ref{eq1}) from experiments (black) and simulations (orange) in the $\Omega_1$-$B_i$ plane with bubble diameter proportional to $\psi$ and $Rm$ shown on the right. The red circle indicates the threshold size $\psi=0.005$ above which cases have prominent SMRI. Straight lines show contours of constant $B_0$. Blue crosses and dots show predictions from global linear analysis that the $m=0$ mode is stable (crosses) or unstable (dots) in a $\Omega(r)$ with $q<2$, which is averaged in the bulk ($-0.25\leq z/H\leq0.25$) of the saturated MHD state with the same $Rm$ and $B_0$ values from simulation.}
\label{fig4}
\end{figure}

Based on the above findings, we define the amplitude $\psi$ of the axisymmetric SMRI as
\begin{align}
&\psi=H(\xi)\xi, \ \text{with} \nonumber\\
&\xi=\langle B_r\rangle/B_i-S\langle B_{r,\rm{ref}}\rangle/B_{i,\rm{ref}}-\left(\beta_0+\beta_1Rm\right).
\label{eq1}
\end{align}
In Eq.~(\ref{eq1}), $H(x)$ is the Heaviside function. $S=1$ for experimental data and $S=0$ for simulation data.
$\beta_0+\beta_1Rm$ is a linear fit to the $\langle B_r\rangle/B_i-S\langle B_{r,\rm{ref}}\rangle/B_{i,\rm{ref}}$ at $Rm\lesssim2.5$ with fixed $B_0$. Here $\beta_0(B_0)$ and $\beta_1(B_0)$ are fitting constants, where $|\beta_0|\lesssim0.002$ for the best fit.
To prevent the influence of fitting uncertainties on the results, we empirically choose $\psi\gtrsim0.005$ as the criterion for prominent SMRI, which is used for solid data points in Fig.~\ref{fig2}. This criterion is consistent with the “knee” position identified in our previous 2D simulations~\cite{WJGEGW20}.
Figure~\ref{fig4} shows that $\psi$ from the experiment and simulation are in good agreement, both showing the typical characteristics of axisymmetric SMRI ($Rm\gtrsim3$ and $B_i\gtrsim2000$~G).
The relatively small $\psi$ in simulation is likely due to its large viscosity (small Re) that reduces the MRI magnitudes. On the other hand, these thresholds are much smaller than the minimum $Rm$ ($Rm\gtrsim9$) and $B_i$ ($B_i\gtrsim5000$~G) required for the onset of axisymmetric SMRI, which are predicted by local Wentzel-Kramers-Brillouin analysis or global linear analysis based on an ideal Couette flow between two infinitely long cylinders~\cite{JGK01,GJ02} (see Fig.~S2 in SM).
We attribute this discrepancy to the imperfect bifurcation nature of the axisymmetric SMRI in our bounded system: under the combined effect of no-slip boundary condition and line-tying effect provided by the conducting end caps, the magnetized Ekman circulation deviates a base flow from the ideal Couette profile, making it easier to excite SMRI.
To confirm this idea, the stability of the $m=0$ mode in a base flow $\Omega(r)$ sampled in the saturated MHD state is tested by global linear analyses, which assume a vertical wavelength of $H$ with insulating radial boundary conditions. Here $\Omega(r)$ is averaged in the bulk region of cases with $B_0\leq0.2$ only, which guarantees that $q<2$. Our global linear analysis confirms that the $m=0$ mode is stable in this $\Omega(r)$ without a magnetic field, as expected.
For $B_0\geq0.3$, local $q>2$ regions shown in Fig.~\ref{fig1} significantly penetrate into the bulk with the increase of $Rm$, making the stability of the $m=0$ mode not solely determined by SMRI and thus no longer an indicator of it~\cite{GGJ12}.
As indicated by the blue crosses and dots in Fig.~\ref{fig4}, the $m=0$ mode at $B_0=0.1$ is stable for all $Rm$ values studied, indicating that the magnetic field is too weak to excite SMRI. For $B_0=0.2$, it becomes unstable for $Rm\gtrsim3.5$, implying the onset of SMRI. This finding is consistent with the emergence of $\psi\gtrsim0.005$ bubbles, thus provides an independent confirmation of the presence of SMRI.

\begin{figure}
\centerline{\includegraphics[width=0.44\textwidth]{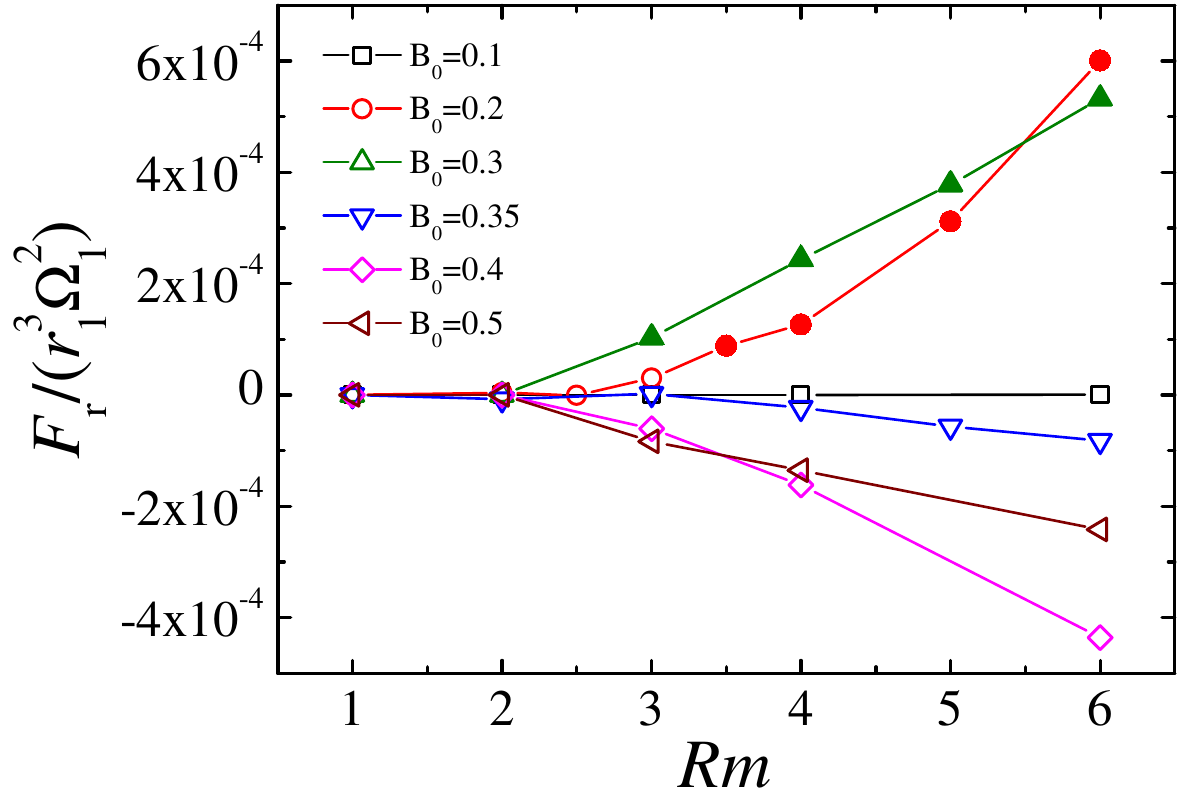}}
\caption{(a) Calculated normalized radial flux of axial angular momentum $F_r/(r_1^3\Omega_1^2)$, as a function of $Rm$ with fixed values of $B_0$ from 3D simulations.
The data are time and volume averages over the bulk region with $1\leq r/r_1\leq3$ and $-0.25\leq z/H\leq0.25$. For each dataset, we have subtracted the linear dependence of $F_r/(r_1^3\Omega_1^2)$ on $Rm$ determined by a linear fit to the data at $Rm\lesssim2.5$. Solid (open) symbols represent cases with (without) significant positive (outward) angular momentum flux, defined as $F_r/(r_1^3\Omega_1^2)\gtrsim10^{-4}$.}
\label{fig5}
\end{figure}

SMRI-induced outward angular momentum flux~\cite{PCP06} can be examined in our system using simulation data.
The radial flux of the axial angular momentum per unit mass is defined as~\cite{BH98,WJGEGJL16}
\begin{align}
F_r(r,z) & =F_H(r,z) + F_M(r,z) \nonumber\\
&=r\left(\delta u_r\delta u_\phi-\nu r\frac{\partial}{\partial r}\delta\Omega-\frac{B_rB_\phi}{\rho\mu_0}\right),
\label{eq2}
\end{align}
where $F_H=r(\delta u_r\delta u_\phi-\nu r\partial_r\delta\Omega)$ is the velocity field contribution and $F_M=-rB_rB_\phi/(\rho\mu_0)$ is the magnetic field contribution.
$\delta u_r(r,z)$, $\delta u_\phi(r,z)$ and $\delta\Omega(r,z)$ are the differences between the $\phi$-averaged radial velocities, azimuthal velocities and angular velocities in the saturated MHD state and the hydrodynamic state.
$B_r(r,z)$ and $B_\phi(r,z)$ are the $\phi$-averaged radial and azimuthal magnetic fields in the saturated MHD state. Equation~(\ref{eq2}) is zero for the hydrodynamic state and thus represents the radial flux of axial angular momentum caused only by axisymmetric MHD effects.
To avoid the interference from the most hydrodynamically unstable ($q>2$) regions adjacent to the ring gaps, we examine Eq.~(\ref{eq2}) in the bulk region.
Figure~\ref{fig5} shows $F_r/(r_1^3\Omega_1^2)$ profiles at different $B_0$, where their linear $Rm$-dependence at small $Rm$ has been subtracted, thereby, all datasets start from zero value and zero slope.
This is to further remove the contribution of the magnetized Ekman circulation to $F_r$, similar to Eq.~(\ref{eq1}).
As shown by the solid symbols, $F_r/(r_1^3\Omega_1^2)$ becomes positive only for $Rm\gtrsim3.5$ with $B_0=0.2$ and $Rm\gtrsim3$ with $B_0=0.3$, a parameter range consistent with that for prominent SMRI (see Fig.~\ref{fig2}b). This finding thereby indicates that the SMRI causes an outward flux of axial angular momentum in the bulk of our system, in line with predictions for accretion disks.
Further analysis reveals that SMRI results in positive values for both $F_H$ and $F_M$, implying that both velocity and magnetic fields contribute outward angular momentum fluxes (see the appendix for more details).
It is also found that $F_H$ is an order of magnitude larger than $F_M$. This is possibly due to the small $Rm$ and large $Re$ regime studied here, where the SMRI can induce only small changes to the imposed magnetic field.

To summarize, we have presented the first convincing experimental evidence for axisymmetric SMRI in a modified Taylor-Couette cell using liquid metal. By measuring the radial magnetic field at the inner cylinder with a proper background subtraction, the $m=0$ SMRI is characterized by the nonlinear increase of the $\langle B_r\rangle/B_i$ as a function of $Rm$ at fixed values of $B_0$. Such increases become prominent only under the conditions of a sufficiently large $Rm$ and an intermediate $B_i$, a feature resembling the typical requirements for SMRI predicted by linear theories.
The experimentally identified SMRI is confirmed independently by 3D numerical simulations in the same geometry. The simulation further reveals that the SMRI produces an outward radial flux of axial angular momentum in the bulk region, an effect similar to that in the accretion disk, further confirming its existence.

Further investigations are needed to fully characterize the reported axisymmetric SMRI and its effects on the liquid-metal flow.
To study detailed temporal evolution, an appropriate way is needed to remove the dynamic radial magnetic field induced by the transient azimuthal currents in conducting end caps and liquid metal, as well as the magnetized Ekman circulation, when $B_i$ is imposed.
Using ultrasonic Doppler velocimetry and Hall probe arrays, different components of velocity and magnetic fields, and their correlations, inside the liquid metal flow will be measured, providing a way to experimentally verify the outward angular momentum transport, as well as the SMRI-induced inward jet in the midplane that has been predicted by recent 2D simulations~\cite{GGJ12,WJGEGW20}.
To overcome the differences in $Re$ between experiments and simulations, simulations with $Re$ closer to experiments should be explored, perhaps with an entropy-viscosity method in SFEMaNS~\cite{GPP11}.
Nonaxisymmetric modes in our system are also of interest, especially for their possible connection to the nonaxisymmetric SMRI in a narrow gap between two infinitely long cylinders predicted recently by a linear theory~\cite{OVBSBLB20}, although the latter has a different geometry. Finally, 2D linear eigenmode calculations of the saturated MHD state in which $\Omega$ depends on both $r$ and $z$ can provide physical insights of the $m=0$ stability beyond 1D global linear analyses to differentiate between SMRI and hydrodynamic instability.

\vskip 0.3cm
Digital data associated with this work are available from DataSpace at Princeton University~\cite{data}.

\vskip 0.3cm
This research was supported by U.S. DOE (Contract No. DE-AC02-09CH11466), NASA (Grant No. NNH15AB25I), NSF (Grant No. AST-2108871), and the Max-Planck-Princeton Center for Plasma Physics (MPPC). E.G. and H.J. acknowledge support by S. Prager and Princeton University. The authors thank Professor Jean-Luc Guermond and Professor Caroline Nore for their permission and assistance in using the SFEMaNS code.

\begin{figure}
\centerline{\includegraphics[width=0.48\textwidth]{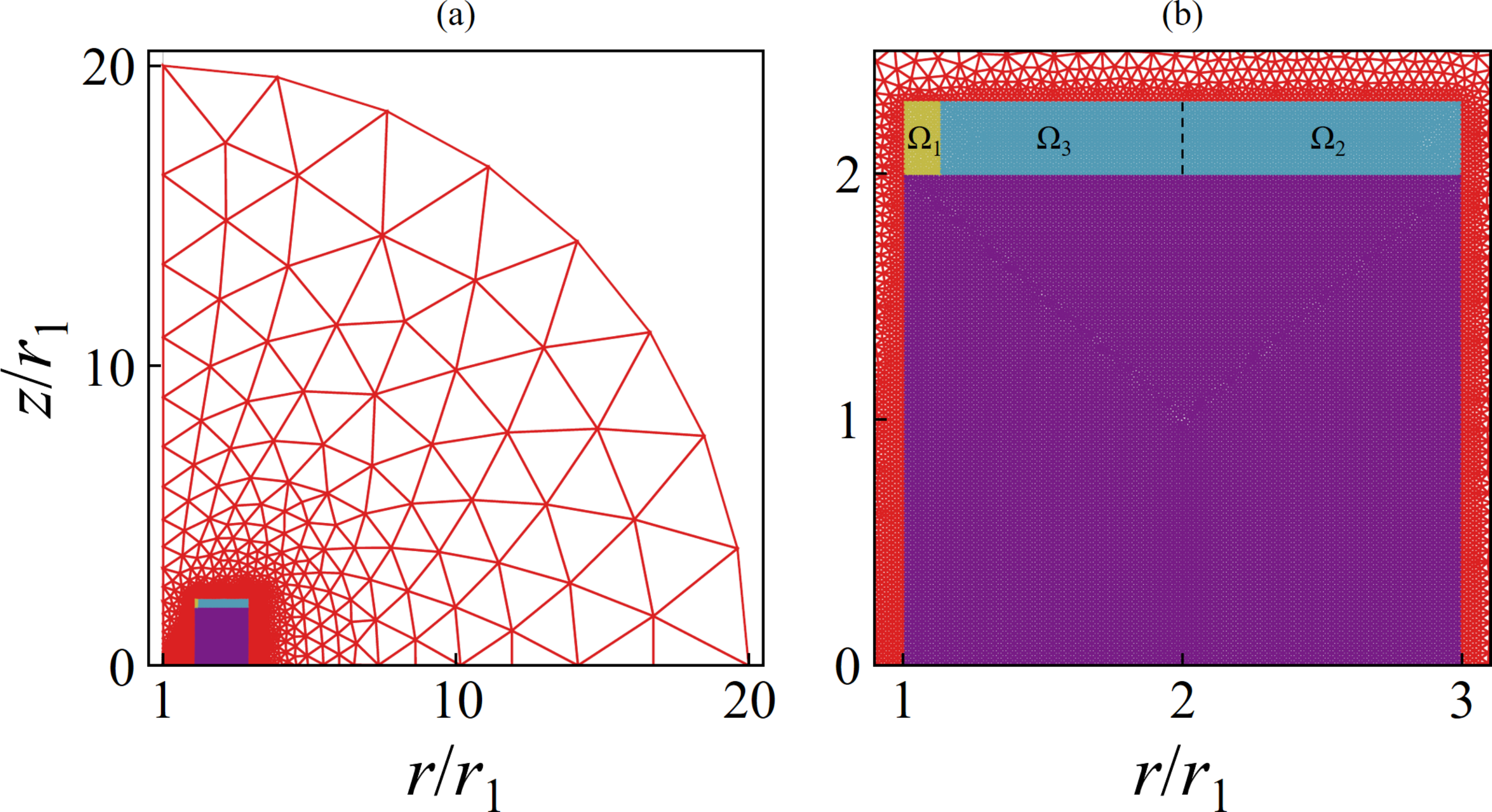}}
\caption{Mesh assignment in a quarter section of the meridional plane (a) and an enlarged portion around the Taylor-Couette cell with rotational speeds of different components marked (b). Colors indicate different domains with purple as the fluid, blue as the copper end cap, yellow as the stainless steel rim of the inner cylinder, and red as the vacuum. The black dashed line in (b) indicates the boundary between the inner and outer rings. This plot is adopted from Ref.~\cite{WJGEGW20}.}
\label{fig6}
\end{figure}

\vskip 0.8cm
\textit{Appendix A: Numerical simulation setup.}---The simulation configuration has been described in detail elsewhere~\cite{WJGEGJL16,CECGGJ19,WJGEGW20}, and only some key points are mentioned here. The SFEMaNS code uses a Fourier spectral method in the azimuth and finite elements in the meridional plane. As in the experiment, a cylindrical coordinate system is adopted in 3D simulations with the origin at the geometric center of the Taylor-Couette cell.
In the simulation, we set $r_2/r_1=3$, $r_3/r_1=2$, $H/r_1=4$ and the radius of the inner cylinder rim $r_\text{rim}/r_1=1.15$. The length, time, velocity, magnetic field and electrical conductivity are normalized, respectively, by $r_1$, $\Omega_1^{-1}$, $\Omega_1r_1$, $\Omega_1r_1\sqrt{\mu_0\rho_g}$ and $\sigma_g$, where $\rho_g$ and $\sigma_g$ are the density and conductivity of galinstan. In order to mimic the experiment, the
entire computation domain is divided into three coupled domains, including a fluid domain for galinstan, a solid domain for end caps and a spherical vacuum domain with a radius of $20r_1$ surrounding them. For all runs, the Reynolds number is fixed at 1000 so the magnetic Prandtl number increases linearly with $Rm$.

Figure~\ref{fig6}a shows the mesh in a quarter of the meridional plane, and meshes with different colors that belong to different domains. The fluid domain has $100\times200$ triangular finite elements in the meridional plane. 32 spectral azimuthal modes are resolved in the azimuthal direction, which is sufficient for the axisymmetric properties studied here, as the volume-averaged energy of all $m\geq1$ modes in the total velocity and magnetic fields  are at least 4 orders of magnitude smaller than the energy of the $m=0$ mode, and the azimuth-averaged velocity and magnetic fields do not change with doubling (64) the number of modes.
Figure~\ref{fig6}b reveals that the solid domain is further divided into two sub-domains with one for the stainless steel inner cylinder rim (yellow) and the other for the copper end cap (blue).
The 3D SFEMaNS code has also been validated against previous experiments investigating the nonaxisymmetric Stewartson-Shercliff layer instability in our system~\cite{CCEGGJ18,CECGGJ19}.

No-slip boundary condition is applied at all fluid-solid interfaces. Conductivity of different components of the end cap is adopted: $\sigma_c=19.4\sigma_g$ for copper and $\sigma_s=0.468\sigma_g$ for stainless steel. Insulating boundary conditions are used for the inner and outer cylinders.
For each run, the simulation starts from a piecewise-solid-body rotation that follows the angular speeds of end-cap components. We first run the hydrodynamic stage for 400 dimensionless time, which is 5 times of the Ekman spin-up time at Re=1000 and thereby is long enough for the flow to reach a statistically steady state~\cite{KJGCS04}.
The axial magnetic field is then imposed and lasts for another 400 dimensionless time, and a saturated MHD state is reached after the first 100 dimensionless time.
All data presented in the main text are based on time averages in the saturated MHD state.

\begin{figure}
\centerline{\includegraphics[width=0.98\textwidth]{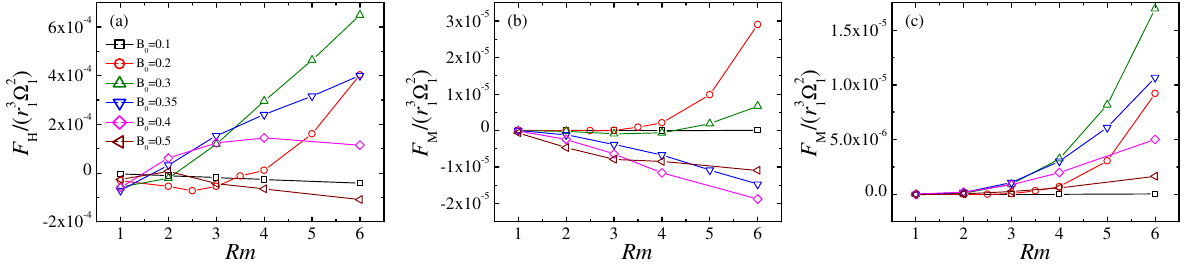}}
\caption{(a) Calculated normalized radial flux of axial angular momentum $F_H/(r_1^3\Omega_1^2)$ contributed by the velocity field, as a function of $Rm$ with fixed values of $B_0$ from simulations.
(b) Corresponding profiles of $F_M/(r_1^3\Omega_1^2)$ contributed by the magnetic field. The data in (a) and (b) are time and volume averages over the bulk region with $1\leq r/r_1\leq3$ and $-0.25\leq z/H\leq0.25$.
(c) The same as (b) but with averages in a local domain close to the inner cylinder with $1\leq r/r_1\leq1.2$ and $-0.25\leq z/H\leq0.25$.}
\label{fig7}
\end{figure}

\vskip 0.8cm
\textit{Appendix B: Analysis of $F_H$ and $F_M$.}---Figure~\ref{fig7}a shows that for calculated $F_H(Rm)/(r_1^3\Omega_1^2)$ as a function of $Rm$, a ``knee'' exists only for high $Rm$ with $0.2\lesssim B_0<0.35$, i.e., $F_H/(r_1^3\Omega_1^2)$ increases faster as $Rm$ increases. This indicates that SMRI prompts the velocity field to generate an outward angular momentum flux, which occupies the middle part of the bulk, and rapidly decays in magnitude and shrinks in area at high $B_0$ (see Fig.~S3 in SM). Similarly, Fig.~\ref{fig7}b shows that a ``knee'' occurs $F_M/(r_1^3\Omega_1^2)$ profiles at high $Rm$ with $0.2\lesssim B_0<0.35$, corresponding to an outward flux contributed by the magnetic field.
It is found that compared with $F_H$, $F_M$ is an order of magnitude smaller and the ``knee'' in it requires higher $Rm$. This is probably due to the small $Rm$ and large $Re$ regime studied here, where the SMRI can induce only small changes to the imposed magnetic field.
When the magnetic field is strong ($B_0\gtrsim0.35$), the bulk-averaged $F_M/(r_1^3\Omega_1^2)$ shown in Fig.~\ref{fig7}b is negative (inward) and decreases with $Rm$. This could be caused by the magnetized Ekman circulation, which originates from regions close to the inner rings and penetrates into the central bulk region at high $B_0$ (see Fig.~S4 in SM).
To avoid this effect, in Fig~\ref{fig7}c the $F_M/(r_1^3\Omega_1^2)$ is averaged only in a local domain close to the inner cylinder, which can well single out the outward flux caused by SMRI without any inward flux.

\onecolumngrid

\twocolumngrid

\clearpage
\onecolumngrid

\section{Supplementary Information}

\begin{figure}[h]
\centerline{\includegraphics[width=0.98\textwidth]{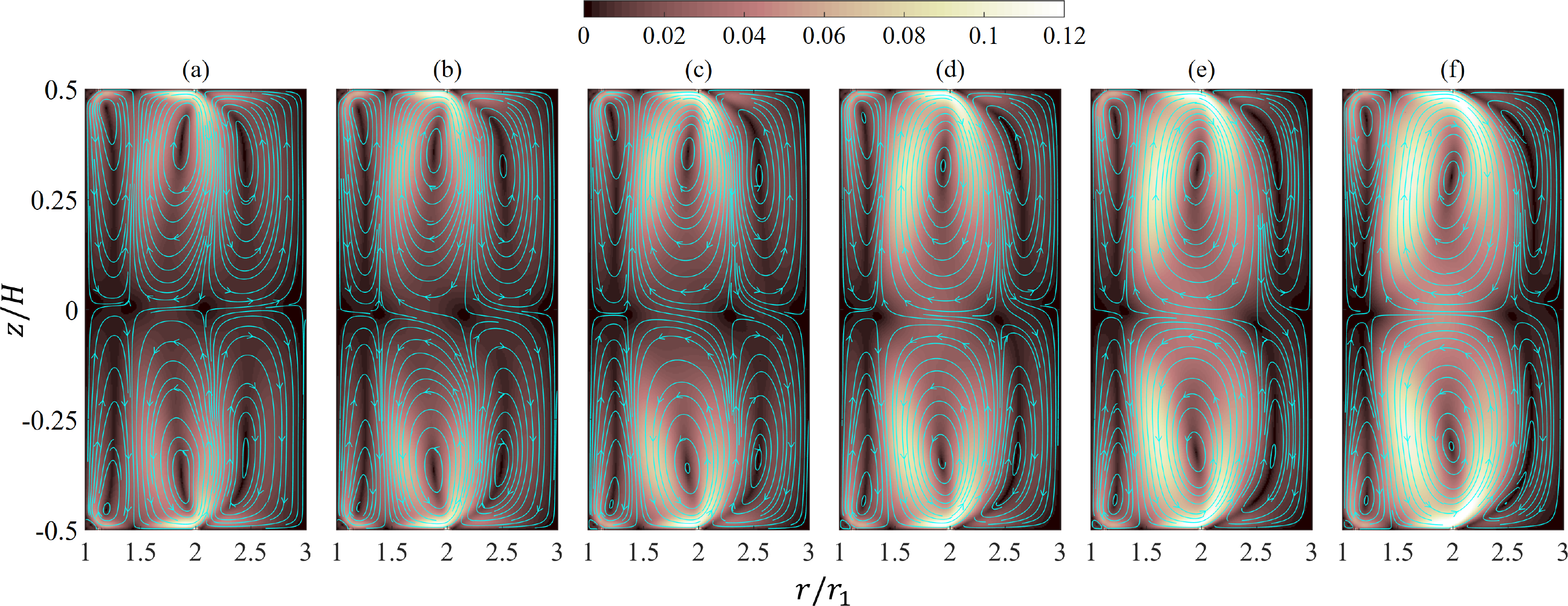}}
\caption{Time and azimuthal average of the normalized meridional speed, $\sqrt{(u_r^2+u_z^2)}/(r_1\Omega_1)$, from three-dimensional simulations. Streamlines for meridional flow are shown as cyan curves.
The data were obtained at $B_0=0.2$ and different values of $Rm$, with (a) $Rm=1$, (b) $Rm=2$, (c) $Rm=3$, (d) $Rm=4$, (e) $Rm=5$, and (f) $Rm=6$. The axisymmetric SMRI is prominent for $Rm\gtrsim3.5$ (see Fig.~2(b) in the main text).}
\label{figS1}
\end{figure}

\begin{figure}[h]
\centerline{\includegraphics[width=0.8\textwidth]{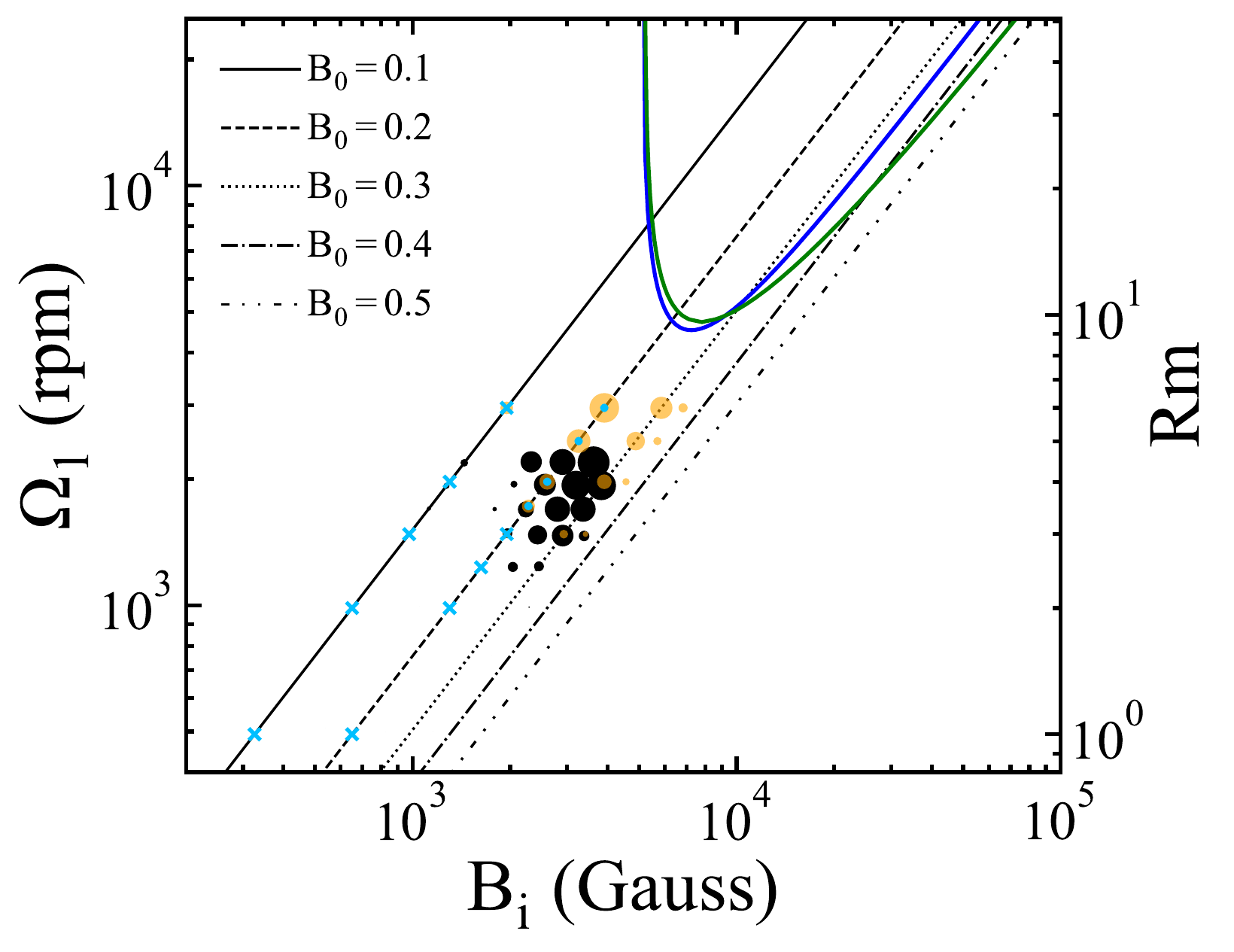}}
\caption{Bubble plot of axisymmetric SMRI amplitude $\psi$ from experiments (black bubbles) and 3D numerical simulations (orange bubbles) in the $\Omega_1$-$B_i$ plane with $Rm$ shown on the right. The data are the same as those in Fig.~4 of the main text. The blue curve represents the boundary of axisymmetric SMRI from Wentzel-Kramers-Brillouin analysis, which assumes a radial wavelength of $2(r_2-r_1)$ and a vertical wavelength of $H$. The axisymmetric SMRI is unstable in the parameter space it encloses. The green curve represents the boundary of axisymmetric SMRI from the global linear analysis, which assumes a vertical wavelength of $H$ with insulating radial boundary conditions.}
\label{figS2}
\end{figure}

\begin{figure}[h]
\centerline{\includegraphics[width=0.98\textwidth]{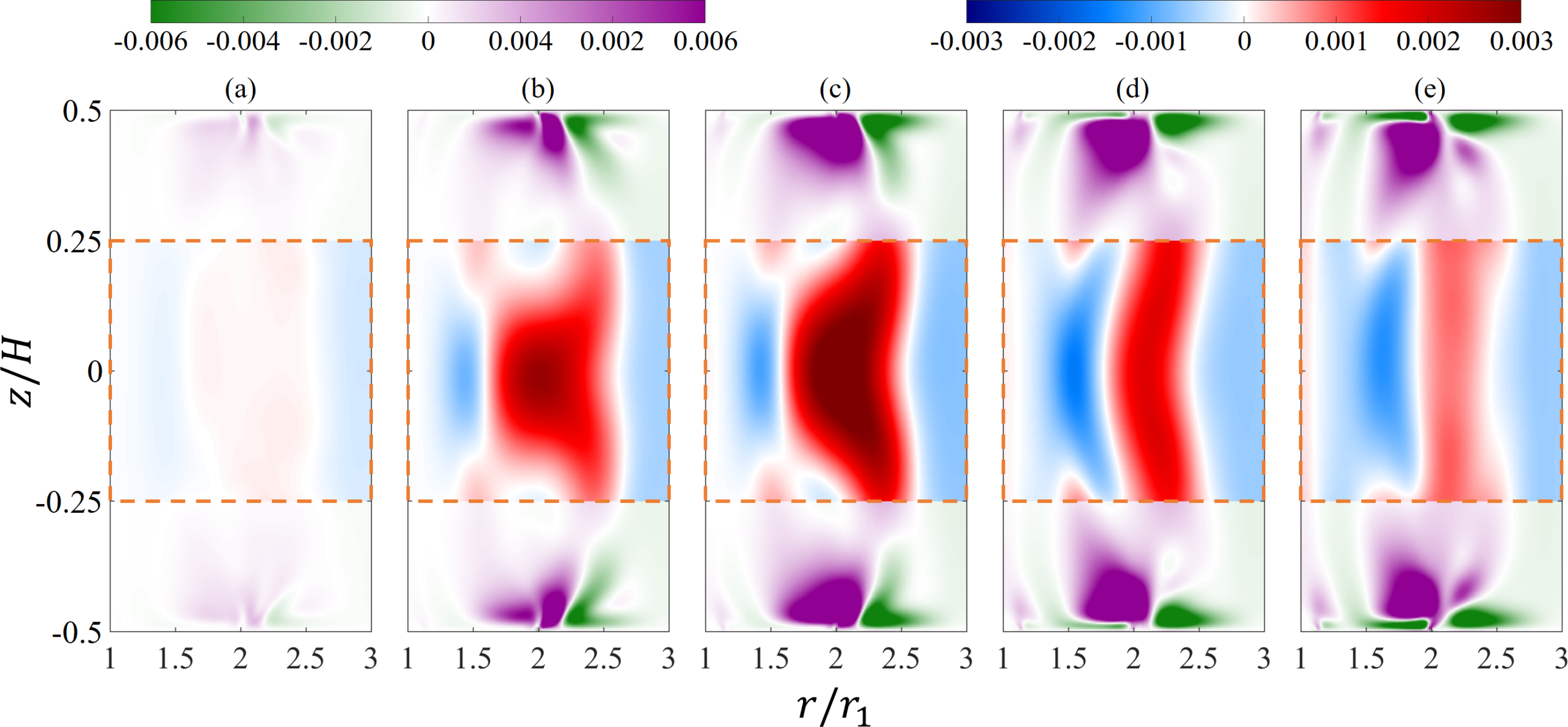}}
\caption{Time- and azimuth-averaged radial flux of normalized axial angular momentum $F_H=r(\delta u_r\delta u_\phi-\nu r\partial_r\delta\Omega)/(r_1^3\Omega_1^2)$ in the meridional plane from three-dimensional simulations. The data were obtained at $Rm=6$ and different values of $B_0$, with (a) $B_0=0.1$, (b) $B_0=0.2$, (c) $B_0=0.3$, (d) $B_0=0.4$, and (e) $B_0=0.5$. The purple-white-green colormap is used for regions close to the end caps ($-0.5\leq z/H\leq-0.25$ and $0.25\leq z/H\leq0.5$). The red-white-blue colormap with a finer range is used for the bulk region ($-0.25\leq z/H\leq0.25$), which is enclosed by the orange dashed box and is the domain where the data in Fig.~7(a) of the main text are volume-averaged.}
\label{figS3}
\end{figure}

\begin{figure}[h]
\centerline{\includegraphics[width=0.98\textwidth]{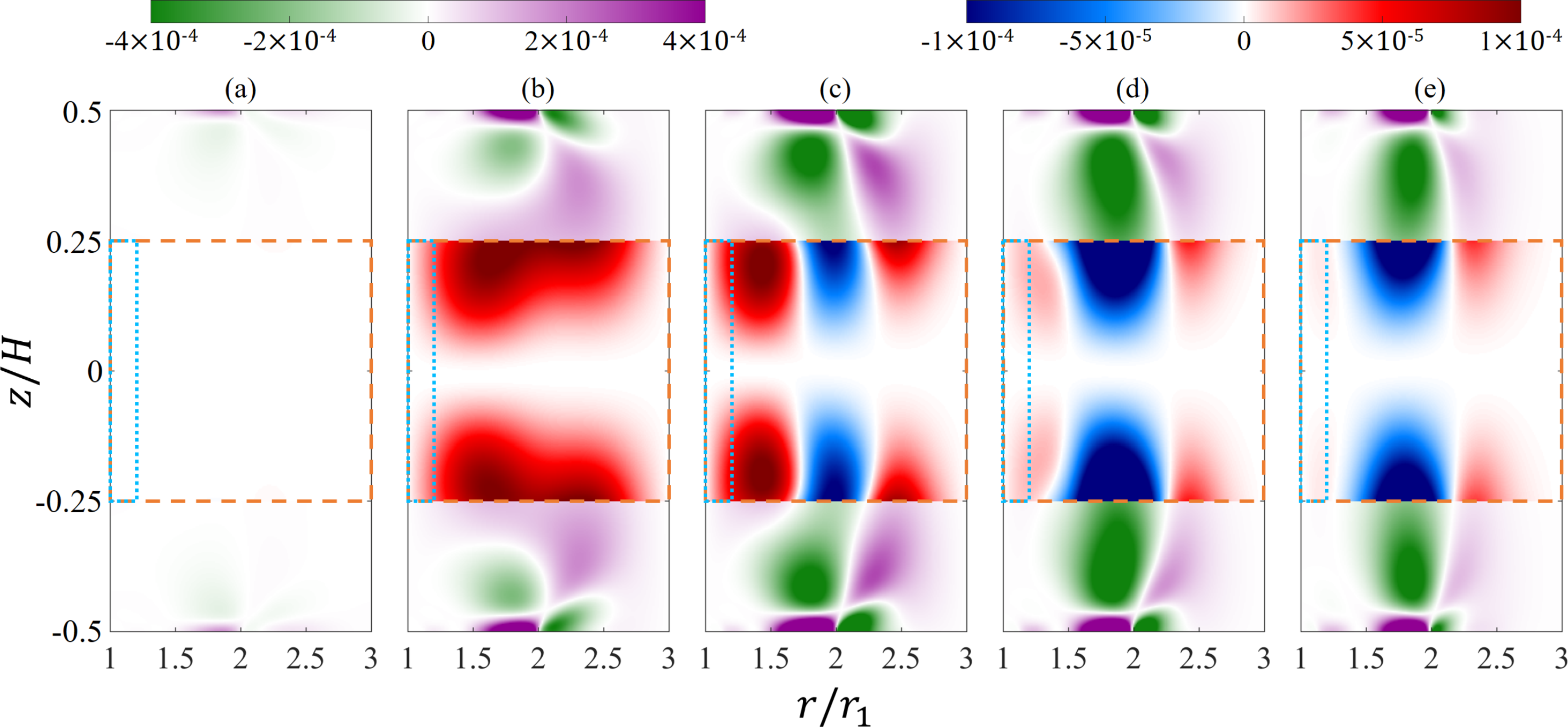}}
\caption{Time- and azimuth-averaged radial flux of normalized axial angular momentum $F_M=-rB_rB_\phi/(\rho\mu_0r_1^3\Omega_1^2)$ in the meridional plane from three-dimensional simulations. The data were obtained at $Rm=6$ and different values of $B_0$, with (a) $B_0=0.1$, (b) $B_0=0.2$, (c) $B_0=0.3$, (d) $B_0=0.4$, and (e) $B_0=0.5$. The purple-white-green colormap is used for regions close to the end caps ($-0.5\leq z/H\leq-0.25$ and $0.25\leq z/H\leq0.5$). The red-white-blue colormap with a finer range is used for the bulk region ($-0.25\leq z/H\leq0.25$). Orange dashed boxes and cyan dotted boxes enclose the domain for volume averaging in Fig.~7(b) and Fig.~7(c) of the main text, respectively.}
\label{figS4}
\end{figure}

\end{document}